\documentclass[
 reprint,
 amsmath,amssymb,
 aps
]{revtex4-2}

\usepackage{graphicx}
\usepackage{dcolumn}
\usepackage{bm}
\usepackage{bbold}
\usepackage{chngcntr}
\usepackage[hidelinks,filecolor=blue,colorlinks=true,linkcolor=blue,citecolor=blue]{hyperref}
\usepackage{float}
\usepackage{dsfont}
\usepackage{braket}
\usepackage{color}

\begin{document}
\preprint{APS/123-QED}

\title{
Thermalization dynamics of macroscopic weakly nonintegrable maps
}

\author{Merab Malishava$^{1,2}$}
\email[Corresponding author:\\]{merabmalishava@gmail.com}
\author{Sergej Flach$^{1,2}$}
\email[Corresponding author:\\]{sergejflach@googlemail.com}
\affiliation{%
\mbox{$^1$Center for Theoretical Physics of Complex Systems, Institute for Basic Science(IBS), Daejeon, Korea, 34126}\\
$^2$Basic Science Program, Korea University of Science and Technology(UST), Daejeon, Korea, 34113
}
\date{\today}
\begin{abstract}
We study thermalization of weakly nonintegrable nonlinear unitary lattice dynamics.
We identify two distinct thermalization regimes close to the integrable limits of either linear dynamics or disconnected lattice dynamics. 
For weak nonlinearity the almost conserved actions correspond to extended observables which are coupled into a long-range network.
For weakly connected lattices the corresponding local observables are coupled into a short-range network. We compute the evolution of the variance $\sigma^2(T)$ of finite time average distributions for extended and local observables. We extract the ergodization time scale $T_E$ which marks the onset of thermalization, and determine the type of network through the 
subsequent decay of $\sigma^2(T)$. We use the complementary analysis of Lyapunov spectra \cite{malishava2022lyapunov} and compare the Lyapunov time $T_{\Lambda}$ with $T_E$. We
characterize the spatial properties of the tangent vector and arrive at a complete classification picture of weakly nonintegrable macroscopic thermalization dynamics.
\end{abstract}

\maketitle

\textbf{Recent studies of the thermalization of macroscopic weakly nonintegrable dynamical lattice systems revealed the existence of two qualitatively different system classes.
Thermalization can be analyzed by testing ergodicity or mixing. From a practical perspective ergodicity and finite time average studies are preferred. 
This calls for a choice of the observables, and we need to keep in mind that the outcome may be choice-dependent.
Two typical observable sets are local observables (LOs) (local norm, charge, energy, etc) and extended observables (EOs) (normal modes).
The LOs become the actions of an integrable limit in the limit of vanishing coupling constant. The EOs correspond to actions in the case of vanishing nonlinearity.
The relation between the thermalization time scales of both types of observables was rarely studied. Moreover, we lack tools to identify the weak nonintegrability class. In this paper, we will utilize a highly efficient numeric scheme – unitary maps – to address the thermalization properties of each choice of observables in close proximity to the corresponding integrable limits.
We demonstrate that a thermalization study using both LOs and EOs allows to unambiguously determine the right system class.
We supplement our studies with additional studies of mixing properties by computing the scaling properties of Lyapunov spectra.
}

\begin{figure}[t!]
    \centering
    \includegraphics[width=\linewidth]{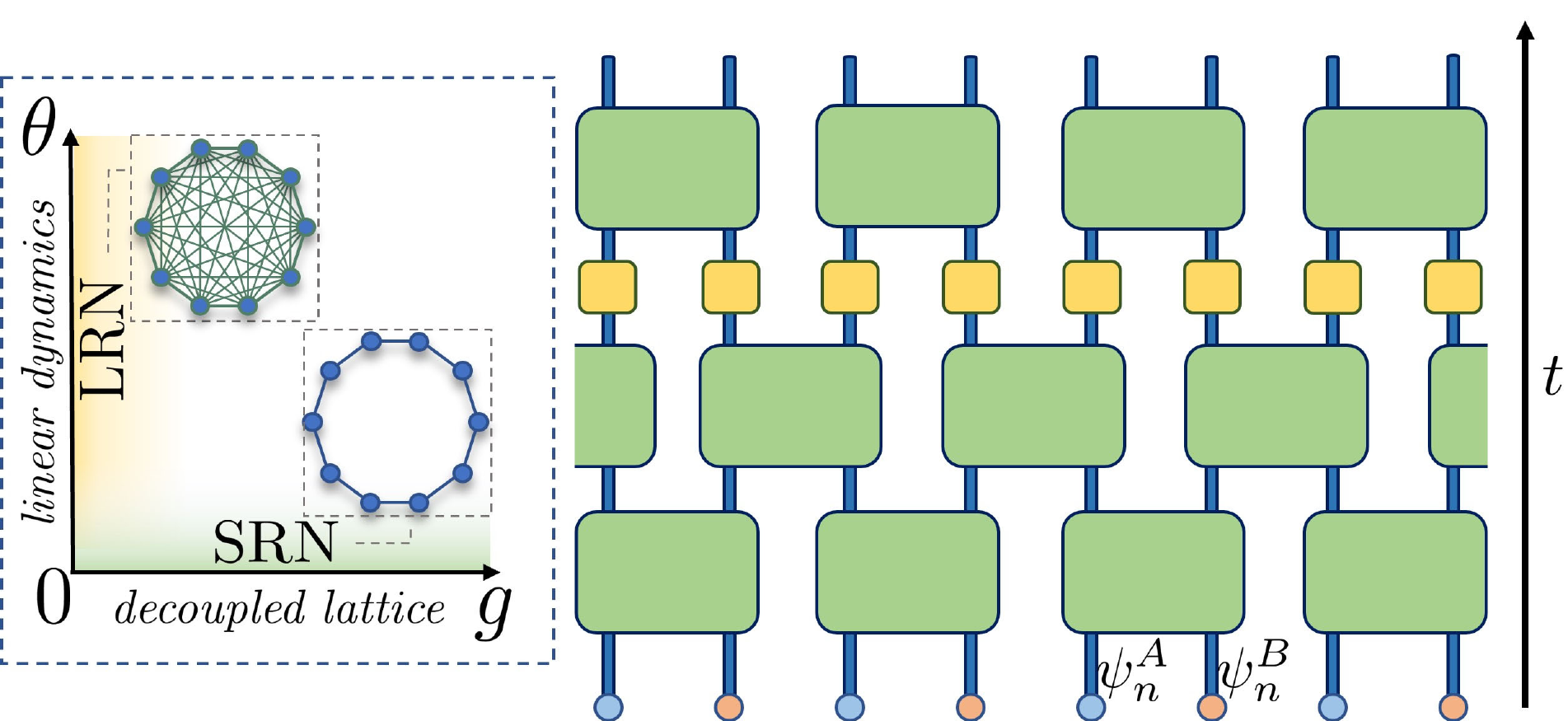}
    \caption{A schematic representation of the unitary circuits map and the parameter space $\{ g, \theta\}$. The black arrow on the right indicates the flow of time. The state is represented by blue and pink dots for $\psi_n^A$ and $\psi_n^B$ respectively, with subsequent applications of unitary matrices $\hat C$  (large green blocks)  parametrized by the angle $\theta$, and local nonlinearity generating maps $\hat G$ (small yellow blocks) parametrized by the nonlinearity strength coefficient $g$. In the parameter space the highlighted areas correspond to the networks of coupled actions induced by respective weak nonintegrable perturbation. Integrable limits are reached for $g=0$ (linear evolution of extended normal modes) or $\theta=0$ (decoupled nonlinear map lattice). Small nonzero $g$ values induce LRNs, small nonzero $\theta$ values induce SRNs. The network images indicate actions (filled circles) coupled due to nonintegrable perturbation (straight lines).
}
    \label{fig1}
\end{figure}

\section{\label{Sec1}Introduction}
The study of thermalization of macroscopically large systems is one of the main goals of statistical mechanics. Thermal behavior assumes the equal a priori probability for the states with equal energy \cite{huang1987statistical}. This notion is tied to the concepts of ergodicity and mixing. Ergodicity assumes the equality between phase space and time averages of observables on one trajectory and is in principle a sufficient condition for a system to demonstrate thermal properties for specifically chosen observables. A stronger property of mixing in addition to ergodicity demands the decay of correlations in time. From the standpoint of evolution in phase space the concept of integrability plays a crucial role. In integrable dynamics trajectories in the phase space are confined to multidimensional tori characterized by sets of actions and angles. Such motion may be ergodic but not mixing. Investigating realistic physical setups with decaying correlations and mixing dynamics begins with deviating from integrability. In one of the first works on the matter Poincare considered a three-body problem which turned out to be impossible to solve analytically, but could be brought to a form of a weakly perturbed integrable system \cite{poincare1890probleme, poincare1893methodes}. 
Later the celebrated theoretical work done in 1954 by Kolmogorov \cite{kolmogorov1954conservation} and extended by Arnold \cite{arnol1963proof} and Moser \cite{moser1962invariant} now known as the KAM theory provided a proof for quasiperiodic dynamics for weakly nonintegrable systems with a finite number of degrees of freedom (DoF). The pioneering numerical studies in weakly perturbed integrable dynamics were performed by Fermi, Pasta, Ulam and Tsingou on a harmonic chain model with weak nonlinear perturbation \cite{fermi1955studies}. Instead of the expected equipartition of energy across the normal modes of a harmonic chain FPUT observed recurrent seemingly quasiperiodic dynamics which is now known as the FPUT paradox \cite{ford1992fermi, campbell2005introduction, gallavotti2007fermi}. The unexpected nonthermal behavior sparked a plethora of research to investigate the absence of thermalization. This led to the discovery of exact energy localizing solutions in terms of solitons \cite{zabusky1965interaction, zabusky1967dynamics}, absence of thermalization in low-dimensional \cite{henon1964applicability} as well as macroscopic non-linear systems \cite{toda1967vibration, tsironis1996slow, matsuyama2015multistage}.  

Most importantly the KAM regime is valid up to a critical strength of the nonintegrable perturbation and replaced by
seemingly homogeneous chaotic dynamics for stronger perturbations. The critical strength is highly sensitive to the number of participating DoF and expected to rapidly deteriorate already for modest system size \cite{wayne1984kam}. This observation is indirectly confirmed by a plethora of computational studies and simply by everyday life, and
serves as the backbone of the validity of the core assumptions of statistical physics. Tuning a macroscopic system close to an integrable limit will nevertheless lead to an increase and divergence of its thermalization time scales. One of the most intriguing questions then is whether the slowing down of thermalization for macroscopic weakly nonintegrable systems 
is unique, or whether there are different classes of such systems beyond the limits set by KAM theory.

To extract ergodic thermalization timescales one usually chooses a set of observables whose dynamics will be followed through the evolution. For weakly non-integrable systems the choice typically falls on the actions of the integrable limit as those are conserved once the perturbation is switched off. Thus, as stated above, FPUT chose the normal modes of a linear system as a set of relevant observables. However, this is not the only possibility. In recent studies variants of the FPUT model were associated with the Toda chain and the equilibration of Toda actions was studied \cite{kurchan2019equilibration}. At the same time, the equipartition of observables weakly correlated to the actions of an integrable limit may show fast equilibration, as studied e.g.  for local energies of an FPUT model \cite{ganapa2020thermalization}. Thus, the choice of observables impacts the thermalization timescale analysis. Moreover, by choosing a specific set of observables the ergodic (but not mixing) dynamics may show thermal behavior \cite{vulpiani2021thermalization, baldovin2021statistical}. In such cases the action-angle dynamics are characterized by incommensurate frequencies and fill the available phase space of multidimensional tori densely, thus displaying ergodic thermal-like behavior without mixing and chaoticity.

In order to study the slowing down of thermalization of a macroscopic system upon approaching an integrable limit, it appears reasonable to choose the actions of the integrable limit as the relevant obserables. Recent studies unraveled that the way these actions are coupled into a dynamical network by the nonintegrable perturbation depends on the perturbation itself
\cite{danieli2019dynamical, mithun2019dynamical, thudiyangal2021fragile}. Translationally invariant linear dynamics results in actions corresponding to extended observables (EO) such as
normal modes. Nonlinear perturbations result from approximations of two-body interactions which are local in real space, but couple all modes with each other due to the fact that the modes themself are extended. The outcome is a class of Long Range Networks (LRN). Typical examples include FPUT, chains of anharmonic oscillators in the limit of weak nonlinearity, Josephson junction arrays in the limit of low energy density, etc. On the other hand, models where the proximity to integrable limit is controlled by a weak lattice coupling constant belong to the class of Short Range Networks (SRN). In this type of scenario the integrable limit is characterized by a set of local observables (LOs), such as a local energy, norm, charge, etc. Models such as coupled anharmonic oscillators in the limit of weak coupling, Josephson junction arrays in the limit of weak coupling, etc. belong to the class of SRN \cite{danieli2019dynamical}. 

In this work, 
we pose the question whether there is a unique protocol which allows to predict the network type of a weakly nonintegrable system which is tuned closer to integrability. In principle the type of network and integrable limit is already encoded in the structure of equations of motion (EoM) which generate the trajectories. We investigate the reverse problem of an observer attempting to deduce the structure of the underlying dynamics while only being able to follow some observables of choice -- a typical laboratory scenario when the precise EoM are unknown.
We intend to analyze the slowing down of ergodicity by following finite time averages of observables, in spite of the sensitivity of thermalization studies to the observable choice. 
We aim to fully characterize the ergodic thermalization dynamics of weakly nonintegrable systems by studying the statistical properties of the full set of finite time averages of local and extended observables. We study ergodic properties of LOs and EOs in both SRN and LRN settings. We demonstrate that  this approach allows for a full and unique characterization of
the thermalization dynamics. When complemented by the computation of the Lyapunov spectrum and the tangent eigenvectors, we conclude that a one to one correspondence leads
to an unambigous detection of either the SRN or LRN networks. We note that ergodic thermalization tests are realizable in experimental setups (up to technical issues), while the Lyapunov spectrum computation appears to have no easy analog on the experimental measurement side.

The computational complexity of the above task is challenging. 
The thermalization times grow quickly as the system approaches the integrable limit, which requires one to perform large timescale computations. Further, in typical time-continuous cases the numerics discretizes time and leads to additional errors which have to be kept at a reasonable level using sophisticated integration schemes. Therefore our strategy is to 
use a novel framework of unitary maps. The key feature of the map dynamics is a discrete-time evolution, which is free from the aforementioned errors. Unitary maps emerged as a concept for efficient quantum computations and quantum algorithms \cite{childs2009universal, santha2008quantum}. Recently they have been also successfully used to simulate classical or nonlinear physical processes: achieve record breaking evolution times in nonlinear wave-packet spreading \cite{vakulchyk2019wave} and use a quantized nonlinearity to observe even slower logarithmic spreading of the wavepackets \cite{mallick2022logarithmic}, Anderson Localization \cite{vakulchyk2017anderson, malishava2020floquet}, soliton dynamics \cite{vakulchyk2018almost}, and topological states of matter \cite{cardano2016statistical} among others. We aim to use the numerical advantages offered by unitary maps to simulate the thermalization dynamics of systems with up to $N = 10^5$ sites (DoF) on timescales of up to $T_{max}=10^9$ time steps in close proximity to integrable limits. 

\section{\label{Sec2}
Model and methods}
\subsection{Setup}
We take inspiration from quantum Unitary Circuits (UCs) systems. Unitary Circuits serve as basis models for quantum computing \cite{aharonov1998quantum, politi2008silica} and as a platform for studies of quantum chaos \cite{mottonen2004quantum} and operator spreading \cite{nahum2018operator}.  Typically a state in quantum UCs is represented in terms of a one dimensional wave function with multiple components per site (usually in terms of qubits) which are then coupled by unitary operations.  

Our classical version of Unitary Circuits is a map acting on a state represented by a one dimensional complex valued vector of size $N$ consisting of $N/2$ unit cells with sites $A$ and $B$:
    \begin{equation}
    \label{eq1}
        \vec \Psi(t) = \{ \psi_n^A(t), \psi_n^B(t) \}_{n=1}^{N/2} \;.
    \end{equation}
 The time evolution of the system is governed by a discrete unitary map consisting of several transformations of $\vec\Psi$:
\begin{eqnarray}
\label{eq2}
\hat U =\sum_{n} \hat G_n \sum_n\hat C_{B, A} \sum_{n} \hat C_{A, B},
\end{eqnarray}
where maps $\hat C_{A, B}$ and $\hat C_{B, A}$ are given by unitary matrices acting on the neighboring sites $(\psi_n^A, \psi_{n}^B)^T$: 
\begin{eqnarray}
\label{eq3}
\sum_{n}\hat C_{A, B}\vec{\Psi}(t) = 
\sum_{n}\begin{pmatrix}
\cos\theta & \sin\theta \\
-\sin\theta & \cos\theta
\end{pmatrix}\begin{pmatrix}
\psi_n^A(t) \\ \psi_n^B(t)
\end{pmatrix}  ,\nonumber\\\nonumber\\
\sum_{n}\hat C_{B, A}\vec{\Psi}(t) = 
\sum_{n}\begin{pmatrix}
\cos\theta & \sin\theta \\
-\sin\theta & \cos\theta
\end{pmatrix}\begin{pmatrix}
\psi_n^B(t) \\ \psi_{n+1}^A(t)
\end{pmatrix}  ,\nonumber\\
\end{eqnarray}
and $\hat G_n$ induces a nonlinearity proportional to the strength coefficient $g$:
\begin{eqnarray}
\label{eq4}
\hat G_n \psi_n^{A,B}(t) = e^{i g |\psi_n^{A,B}(t)|^2} \psi_n^{A,B}(t).
\end{eqnarray}
The local transformations $\hat C$ can be in general represented as arbitrary $2\times 2$ unitary matrices. Our particular choice is parametrized with a single angle $\theta$ which plays
the role of a hopping parameter strength in Hamiltonian systems. At the same time the nonlinearity $g |\psi_n|^2$ is an analog of an effective mean-field potential. The maps of this type can be constructed experimentally with periodic application of magnetic field pulses \cite{CoinOperImpl,Impl1,di2004cavity,Impl3}.

The evolution (Eq.~\eqref{eq2}) results in the following equations of motion:
\begin{widetext}
\begin{eqnarray}
&& \psi_n^{A}(t + 1) = e^{i g |\varphi_n^A(t)|^2} \varphi_n^A(t)\;\;\;,\;\;\; \varphi_n^A(t)=
\big[\cos^2\theta\psi_n^A(t)- \cos\theta\sin\theta\psi_{n-1}^B(t) + \sin^2\theta\psi^A_{n +1}(t) + \cos\theta\sin\theta\psi_n^B(t)\big] , \nonumber \\
\nonumber\\
&& \psi_n^{B}(t + 1) = e^{i g |\varphi_n^B(t)|^2} \varphi_n^B(t) \;\;\;,\;\;\; \varphi_n^B(t) =
\big[\sin^2\theta\psi_{n - 1}^B(t)- \cos\theta\sin\theta\psi_{n}^A(t) + \cos^2\theta\psi^B_{n}(t) + \cos\theta\sin\theta\psi_{n + 1}^A(t)\big],
\label{eomsNonLinear}
\end{eqnarray}
\end{widetext}
where $\varphi^{A,B}_n(t)$ are the components of state vector $\vec\Psi$ after the application of mixing maps $\hat C$ (see Fig.~\ref{fig1}).

\subsection{Integrable limits}

We consider two integrable limits in the parameter space $\lbrace g,\; \theta \rbrace$ of the classical Unitary Circuits map: vanishing coupling: $\{\theta = 0, \; g\neq 0\}$, and vanishing nonlinearity: $\{g = 0, \; \theta \neq 0\}$ (see Fig.~\ref{fig1}). In each case the integrable limit is characterized by a corresponding set of $N$ observables (actions) which are decoupled and thus are conserved in time during the evolution. In what follows we will investigate each of the integrable limits in detail by deviating from it slightly and thus inducing a coupling between the actions. 

\subsubsection{Short-range network}

In the limit of vanishing coupling $\theta = 0$ the local transformations $\hat C$ turn into identity matrices resulting in a trivial evolution of the state components with just an accumulation of phase (see Eq.~\eqref{eomsNonLinear} for details):
\begin{eqnarray}
\label{eq6}
    \psi_n^{A,B}(t) = e^{i g |\psi_n^{A,B}(t)|^2 t}\psi_n^{A,B}(0).
\end{eqnarray}
The amplitudes $|\psi^{A,B}_n|$ are time independent and correspond to the actions of the integrable limit. Introducing a small but nonzero value of $\theta \neq 0$ results in a  coupling of the actions. Approximating Eq.$\eqref{eomsNonLinear}$ for small values of $\theta$ we obtain: 
\begin{eqnarray}
&& \psi_n^{A}(t + 1) = e^{i g |\varphi^A_n(t)|^2} \varphi^A_n(t), \nonumber \\
&& \varphi^A_n(t) =
\big[\psi_n^A(t)- \theta ( \psi_{n-1}^B(t) - \psi_n^B(t) ) \big], \nonumber \\
\\
&& \psi_n^{B}(t + 1) = e^{i g |\varphi^B_n(t)|^2} \varphi^B_n(t), \nonumber
\\
&& \varphi^B_n(t) = \big[\psi^B_{n}(t) + \theta ( \psi_{n + 1}^A(t)- \psi_{n}^A(t) ) ]. \nonumber
\label{eomsNonLinear-2}
\end{eqnarray}
These equations of motion couple the actions through nearest neighbor terms and fall under the definition of a short range network (see Fig.~\ref{fig1}). 
\\
\\
\subsubsection{Long-range network}

In the linear case $g = 0$ the evolution of the state vector $\vec\Psi(t)$ can be determined exactly from standard ansatz $\left(\psi_n^A(t), \psi_n^B(t)\right)^T = e^{-i(\omega_k t - k n)}\left(\psi^A_k, \psi^B_k\right)^T$. The eigenfrequencies $\omega_k$ obey the following dispersion relation, which can be determined from Eq.~\eqref{eomsNonLinear}:
\begin{eqnarray}
\label{eq5}
\omega(k) = \pm\arccos\left( \cos^2\theta + \sin^2\theta\cos k\right),
\end{eqnarray}
with two dispersive bands $\omega_k^\alpha$ ($\alpha = 1,2$) and corresponding normal modes $\vec\Psi_k^{\alpha} = \sum_n e^{i k n}\psi_k^{\alpha, p}$ ($p = A, B$) which form a complete set.


Generally, a state vector $\vec \Psi(t)$ may be decomposed in terms of normal modes of a linear system:
\begin{eqnarray}
\label{eq7}
\vec\Psi(t)=\sum_k c_k^\alpha(t)\vec\Psi_k^\alpha.
\end{eqnarray}

In the linear case $g=0$ the evolution of the coefficients $c_k^\alpha (t)$ is given by a phase rotation $e^{i\omega_k^\alpha t}$; the absolute values $|c_k^\alpha|$ are conserved in time, they are the actions of the integrable limit. 
Introducing a small but nonzero value of $g \neq 0$ results in a  coupling of the actions. Approximating Eq.$\eqref{eomsNonLinear}$ for small values of $g$ we obtain: 
\begin{widetext}
\begin{eqnarray}
   \label{eq8}
   c_k^\alpha(t+1) = e^{i \omega_k} c_k^\alpha(t) + 
  \frac{i g}{N}\sum_{\substack{\alpha_1,\alpha_2,\alpha_3 \\ k_1,k_2,k_3}}e^{i (\omega_{k_1}^{\alpha_1} + \omega_{k_2}^{\alpha_2} - \omega_{k_3}^{\alpha_3})}
    I_{k,k_1,k_2,k_3}^{\alpha,\alpha_1,\alpha_2,\alpha_3}c^{\alpha_1}_{k_1}(t)c^{\alpha_2}_{k_2}(t)\left(c^{\alpha_3}_{k_3}(t)\right)^* ,
\end{eqnarray}

\begin{eqnarray}
\label{eq9}
   I_{k,k_1,k_2,k_3}^{\alpha,\alpha_1,\alpha_2,\alpha_3} = \delta_{k_1+k_2-k_3-k,0}\sum_p \psi_{k_1}^{\alpha_1, p}\psi_{k_2}^{\alpha_2,p}(\psi_{k_3}^{\alpha_3, p})^*(\psi_k^{\alpha,p})^*.
\end{eqnarray}
\end{widetext}
All $c_k^{\alpha}$ are intercoupled according to the second term in Eq.~\eqref{eq8}. For each action $c_k^\alpha$ the number of elements in the sum is proportional to $N^2$ due to the constraints enforced by the overlap integrals in Eq.~\eqref{eq9}. This case falls under our definition of a long range network.

\subsection{\label{Sec3}Finite time averages}

Our goal is to study the ergodization of local and extended observables in two distinct integrable limits. Ergodization is a process of time averages of observables approaching their phase space averages. In line with this definition we will define sets of finite time averages of observables and study the statistical properties of these sets, in particular their variance.
In both SRN and LRN regimes we study the statistical properties of sets of finite time averages of observables. For a time-dependent observable $o(t)$ we define a finite time average as:
\begin{eqnarray}
    \label{eq10}
    \overline{o}_T = \frac{1}{T}\sum_{t=0}^{t=T} o(t)
\end{eqnarray}
We construct a set of finite time averages $\lbrace \overline{o}_T \rbrace_M$ by following $M$ trajectories. This set is characterized by a distribution with probability density function $\rho (\bar o_T)$. From the ergodization hypothesis we expect each $\bar o_{T \rightarrow \infty} = \langle o \rangle$, where $\langle o \rangle$ is an average taken over the phase space. The distribution $\rho$ is therefore expected to peak around the phase space average, reducing its variance to zero and thus approaching a delta-function for infinite averaging times: $\rho (\bar o_{T\rightarrow \infty}) = \delta( \bar o_{T\rightarrow \infty} - \langle o \rangle)$. We study the convergence by following the variance $\sigma^2(T)$ of the distribution $\rho (\bar o_T)$.
We perform the analysis for two distinct observables -- local observables $|\psi_n|$, and extended observables $|c_k^\alpha|$ which are the amplitudes of normal modes coefficients.

\subsection{Lyapunov times and tangent vectors\label{sec_model_lyapunovTimesAndTangentVectors}}
To characterize chaoticity related timescales of non-integrable dynamics we compute the full spectrum of Lyapunov characteristic exponents (LCEs). An advantage of LCE spectra is that they are independent of any coordinate basis choice.  Nonzero Lyapunov spectra necessarily imply chaotic \textit{mixing} dynamics (while the converse is not necessarily the case). On the downside we note that while the largest LCE can be experimentally assessed, the entire Lyapunov spectrum appears to be not experimentally measurable.

We compute the full LCE spectra using methods described in \cite{benettin1980lyapunov} in both SRN and LRN limits upon variation of $\theta$ and $g$ respectively. 
We evolve a set of deviation vectors in the phase space and extract the Lyapunov exponents corresponding to their growth, see Appendix \ref{appendix:Lyapunov times} for more details. The number of Lyapunov characteristic exponents equals the dimensionality of the phase space, in our case $4N$. Lyapunov exponents come ordered  from largest to smallest value upon incrementing the index $i$. Due to the symplectic nature of the unitary circuits map the spectrum is symmetric with LCEs coming in pairs $\Lambda_i = -\Lambda_{2N - i + 1}$. Norm conservation ensures two vanishing LCEs $\Lambda_N = \Lambda_{N+1} = 0$. 

Without loss of generality we consider only positive LCEs. We renormalize Lyapunov spectra $\overline{\Lambda}_i=\Lambda_i/\Lambda_{max}$ and rescale the index index $\rho = i/2N$ so that all positive LCEs $\overline{ \Lambda}(\rho)$ correspond to $\rho \in [0,1]$.

An in depth investigation of Lyapunov spectra scaling of weakly nonintegrable dynamics has been performed in Ref.~ \cite{malishava2022lyapunov}.

Further, we perform a characterization of the tangent vector corresponding to the largest LCE (for definition of tangent vectors see Appendix \ref{appendix:Lyapunov times}). At any given time the normalized tangent vector points in the direction of the strongest chaotization. We compute the time average of the participation number $\overline{PN}$ of the normalized tangent vector $\vec w(t)$:
\begin{equation}
   PN(t)=\frac{1}{\sum_i |w_i(t)|^4}.
\end{equation}
The tangent vectors depend on the coordinate basis choice. Once the basis is fixed, a delocalized tangent vector results in $PN \sim N$ while a localized
tangent vector yields a system size independent value of $PN$.

\subsection{Evolution}
We perform the evolution of the state vector $\vec \Psi(t)$ in coordinate space using the unitary map defined in Eq.\eqref{eq2}. 
Before measuring the observables the system is prerun to ensure thermalization. 
The initial conditions for each component of the state vector are set as $\psi_n^{A,B} = r_n e^{i\gamma_n}$. For each of the $M$ initial conditions (trajectories) we generate the amplitudes $r_n$ as uncorrelated random numbers to be distributed according to an exponential distribution with probability density function $f(x)\sim x e^{-x^2}$ in accord with the Gibbs distribution for the norm densities. We then renormalize the state vector such that the norm density is set to unity. The phases $\gamma_n$ are distributed uniformly on the interval $[0, \; 2\pi]$.

\subsection{Observables}

    \textbf{LOs:} Due to the translational invariance in the system, we assume the local observables (LOs) to be statistically identical and independent. This leads to the possibility of obtaining a set of finite time averages of LOs from a single trajectory. This way an output of a single run will generate the variance $\sigma^2(T)$ of a set of finite time averages of $N$ observables.
    
    \textbf{EOs:} In contrast to LOs the normal mode coefficients $c_k^\alpha$ - which are the extended observables (EOs) - are not statistically identical. Thus we choose a specific value of the wave vector $k = \pi/2$ and perform $M$ trajectory runs. We extract the time average $\overline{c}_k^m(T)$ to obtain the set of $M$ finite time averages $\lbrace \overline{c}_k^m (T) \rbrace_{m=1}^M$ whose variance $\sigma^2(T)$ is then computed.
    
\subsection{Time scales}

For the sake of clarity we briefly list and define the time scales involved in our studies. The ergodization time $T_E$ is the time scale up to which the
actions which turn integrals of motion at the very integrable limit stay essentially constant for a weakly nonintegrable system. It follows that $T_E$ must diverge upon approaching the very integrable limit. We study systems with short range coupling in real space, which co-exists with any of the nonintegrable network ranges (short and long), as the latter are defined in the corresponding action space of actions which turn integrals of motion at the very integrable limit. Therefore, regardless of the type of nonintegrable network, once the time scale $T > T_E$ is reached, diffusion in real space will set in. 
A second diffusion time $T_D \sim N^2$ marks the time at which the diffusion in real space reaches the boundaries of the finite system. Note that for an infinite system size $T_D$ diverges for any weakly nonintegrable system which still has a finite (though potentially very large) ergodization time $T_E$.
Finally, we will compare the above time scales with the Lyapunov time $T_{\Lambda} = 1/\Lambda_{max}$ which is given by the inverse of the largest Lyapunov exponent $\Lambda_{max}$. In all cases $T_{\Lambda} \leq T_E$ which is an expected result -  there can be no ergodization and thermalization before any chaos sets in. However, we find that the ratio $T_E/T_{\Lambda}$ diverges much faster with increasing $T_E$ for SRNs, while stays almost constant for LRNs.

\section{\label{Sec4}Expected Results}

\subsection{Short-range network}
\begin{figure}[t!]
\includegraphics[width=\linewidth]{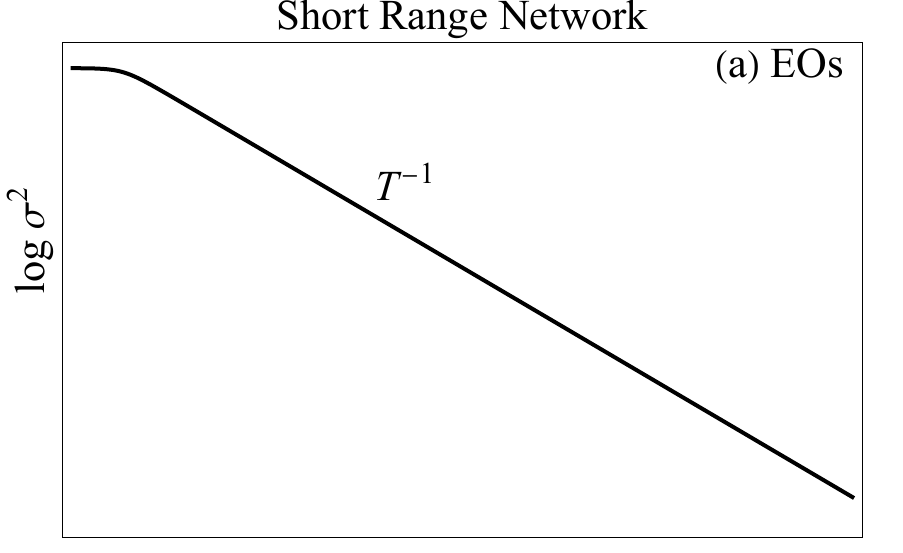}
\includegraphics[width=\linewidth]{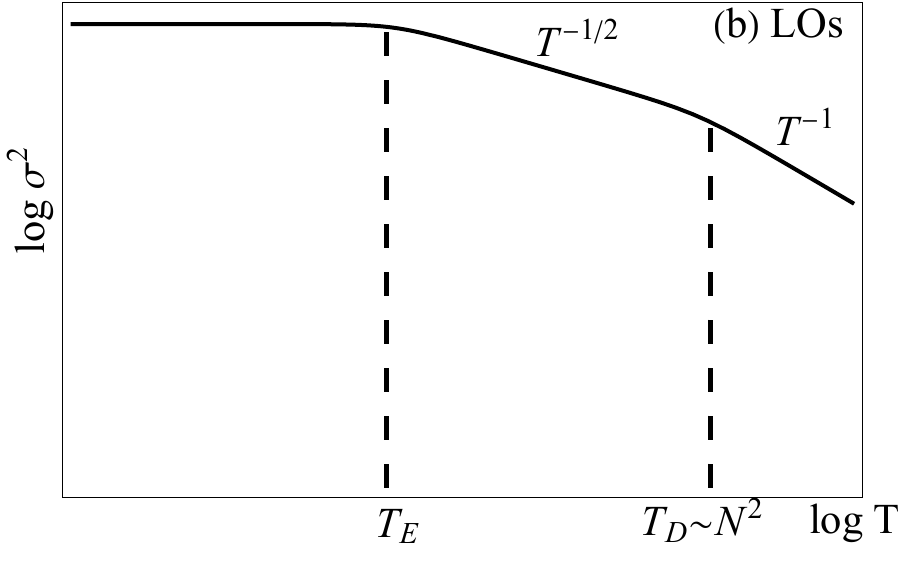}
\caption{\label{fig2} Schematic representation of the expected behavior of the variance $\sigma^2(T)$ in the SRN case in log-log scale. 
(a) The variance of finite time averages of EOs is expected to show the decay $\sigma^2(T) \sim T^{-1}$ according to the central limit theorem as they behave as independent uncorrelated observables. 
(b) The variance of finite time averages of LOs stays constant until $T_E$ (which will diverge at the integrable limit). For $T>T_E$ we expect $\sigma^2\sim T^{-1/2}$ due to diffusive propagation of resonances in the system. Once diffusion reaches the system boundary at $T_{D}\sim N^2$ we expect $\sigma^2 \sim T^{-1}$. 
}
\end{figure}
\begin{figure}[t!]
\includegraphics[width=\linewidth]{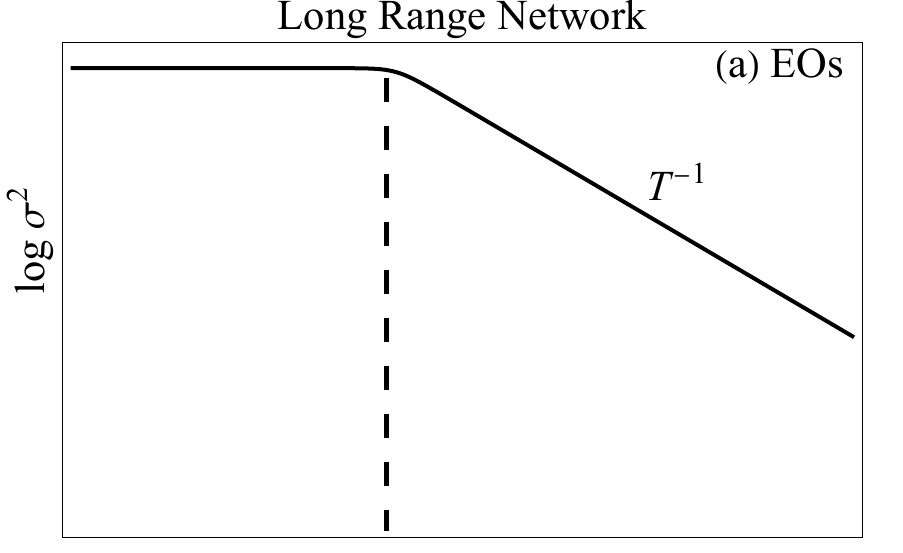}
\includegraphics[width=\linewidth]{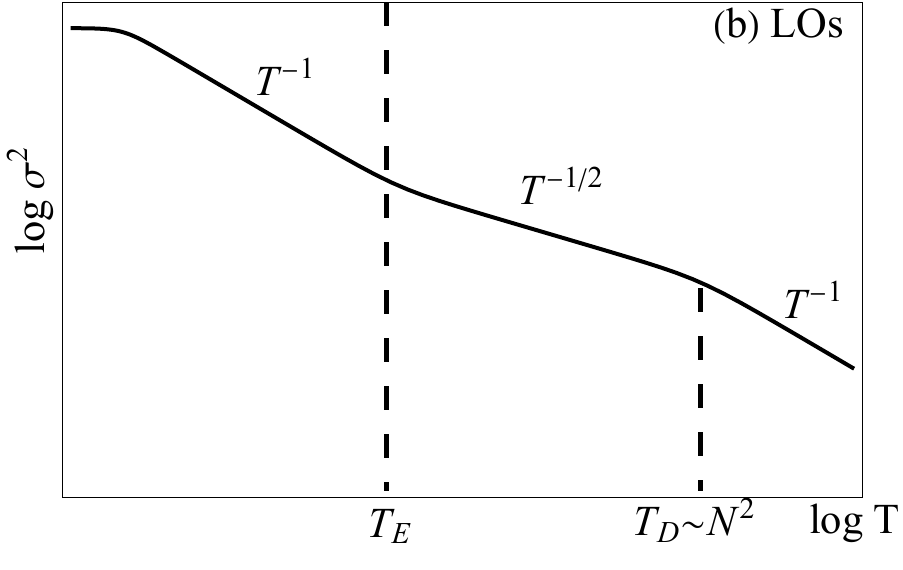}
\caption{\label{fig3} Schematic representation of the expected behavior of the variance $\sigma^2(T)$ in the LRN case in log-log scale. 
(a) The variance of finite time averages of EOs stays constant up to $T_E$. 
For $T>T_E$ we expect $\sigma^2 \sim T^{-1}$. 
(b)
The variance of finite time averages of LOs is expected to show the decay $\sim T^{-1}$ according to the central limit theorem as they behave as independent uncorrelated observables. 
Past the ergodization time $T>T_E$ we expect a transition to $\sigma^2 \sim T^{-1/2}$ due to normal diffusion in real space up to $T_D\sim N^2$.
For large times $T > T_D$ a return to truly uncorrelated decay $\sigma^2 \sim T^{-1}$ is expected. 
}
\end{figure}
In the integrable limit $\theta = 0$ the LOs are decoupled and the variance of the set of finite time averages of LOs will stay constant over time $\sigma^2(T) = const$.
Once the small deviation $\theta \neq 0$ has been introduced the LOs are coupled and the dynamics shows nonintegrable behavior. The weak coupling of LOs will manifest in nearly frozen actions with rare resonant spots in the system, where chaotic dynamics takes place \cite{mithun2019dynamical}. 
The mean distance between the chaotic spots grows upon approaching the integrable limit. The strength of chaotic dynamics (largest Lyapunov exponent) diminishes upon approaching the integrable limit. This interplay between chaotic and non-chaotic parts of the system will result in the ergodization time $T_E$ - a time scale on which the chaotic spots diffuse over a distance of the order of the average spacing between the resonances. It follows that $\sigma^2(T < T_E)$ stays approximately constant up to $T_E$. That time scale $T_E$ will diverge upon approaching the integrable limit.
At finite distance from the integrable limit the resonances continue to diffuse through the system resulting in $\sigma^2(T) \sim T^{-1/2}$ for $T_E < T < T_D$ \cite{thudiyangal2021fragile}.
Once the excitations diffuse across the entire system all correlations vanish and we expect $\sigma^2(T) \sim T^{-1}$ due to finite size effects. The time scale of the transition from $\sigma^2(T) \sim T^{-1/2}$ to $\sigma^2(T) \sim T^{-1}$ is denoted as $T_D \sim N^2$ \cite{thudiyangal2021fragile}.  The schematic representation of expected behavior of $\sigma^2(T)$ is presented in Fig.~\ref{fig2}.

The EOs are not conserved even at the very integrable limit. They will show fast fluctuations and quick pseudo-thermalization in analogy with Ref.~\cite{vulpiani2021thermalization}. 
Thus we expect an immediate $\sigma^2(T) \sim T^{-1}$ decay starting from the shortest time scales, see Fig.~\ref{fig2}.

\subsection{Long-range network}

In the integrable limit $g=0$ the system dynamics is linear, and we expect $\sigma^2(T) = const$ for EOs. Upon the deviation from the limit the variance is expected to decay $\sigma(T)^2 \sim T^{-1}$ after some ergodization time $T_E$ required for spread of chaos into the network (see Fig.~\ref{fig3}). We expect time $T_E$ to be of the order of Lyapunov time $T_\Lambda$ as there appears to be no other time scale governing the dynamics, see Fig.~\ref{fig3}.

LOs will show a more involved outcome. First, even at the integrable limit they are not conserved, and will show fast fluctuation and quick pseudo-thermalization
in analogy with Ref.~\cite{vulpiani2021thermalization}.  Thus we expect an immediate $\sigma^2(T) \sim T^{-1}$ decay starting from the shortest time scales.
However, this only holds up to $T_E$ if the system is large enough. Indeed, the EOs are preserved up to $T_E$ and correspond to ballistically propagating
modes (waves) in real space having a largest finite group velocity $v_g$. If the system size $N \gg v_g T_E$ the modes will start to interact and ballistic
propagation is replaced by diffusive propagation in real space for $T > T_E$. Correspondingly the LO dynamics results in a crossover from $\sigma^2(T) \sim T^{-1}$ to
$\sigma^2(T) \sim T^{-1/2}$ for $T > T_E$. At a time scale $T_D \sim N^2$ the diffusion reaches the system boundaries, and the LO dynamics crosses over back to a
final asymptotic $\sigma^2(T) \sim T^{-1}$ decay. 
%
We show the expected behavior in Fig.~\ref{fig3}.

\section{\label{Sec5} Numerical Results}

\subsection{Short-range network}
In the SRN case we fix the nonlinearity strength $g = 1$ without loss of generality. The system size $N=10^4$. At $\theta = 0$ the LOs are frozen -- the system is at the integrable limit. Upon increasing the parameter $\theta$, the LOs get intercoupled with nearest neighbors into a SRN. We study the statistics of local and extended observables in the SRN in close proximity to the corresponding integrable limit.

\subsubsection{Extended observables}
In 
Fig.~\ref{fig4}(a) we plot the variance $\sigma^2(T)$ for finite time averages of EOs for various values of $\theta$ approaching zero.  All curves show the same result - an immediate decay $\sigma^2 (T) \sim T^{-1}$ from the shortest averaging times on. This holds even in the integrable limit itself for $\theta = 0$. Similar results have been shown by \cite{vulpiani2021thermalization} -- due to the central limit theorem a random transformation of the set of action variables shows statistical properties similar to those of a mixing system when indeed it is not. Such a measurement does not necessarily imply true ergodicity, as follows from what comes next. It also does not allow for a measurement of the large but finite ergodization time scale $T_E$. The observed thermalization is in good agreement with the prediction in Fig.~\ref{fig2}(a).

\subsubsection{Local observables}
\begin{figure}[t]
    \centering
    \includegraphics[width=\linewidth]{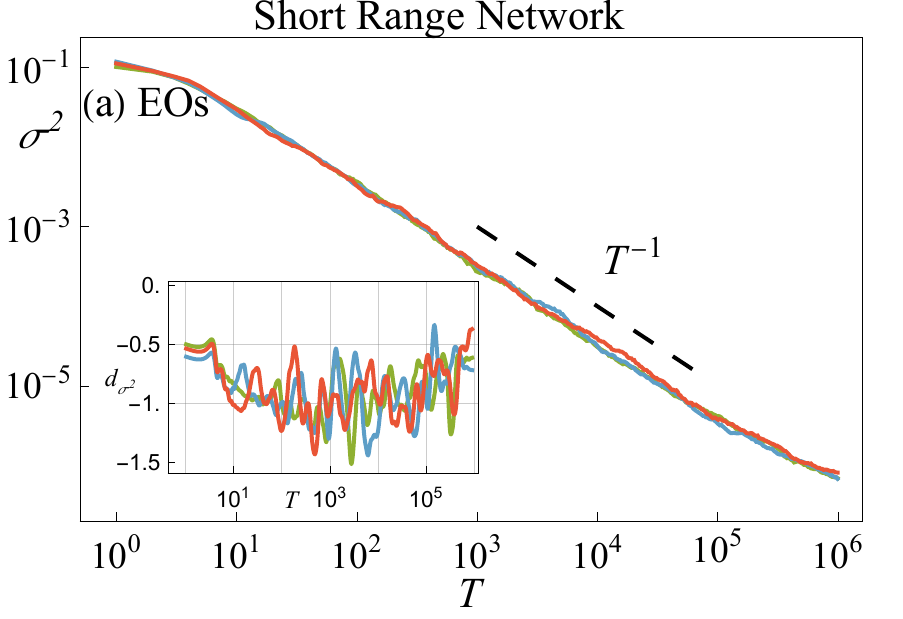}
    \includegraphics[width=\linewidth]{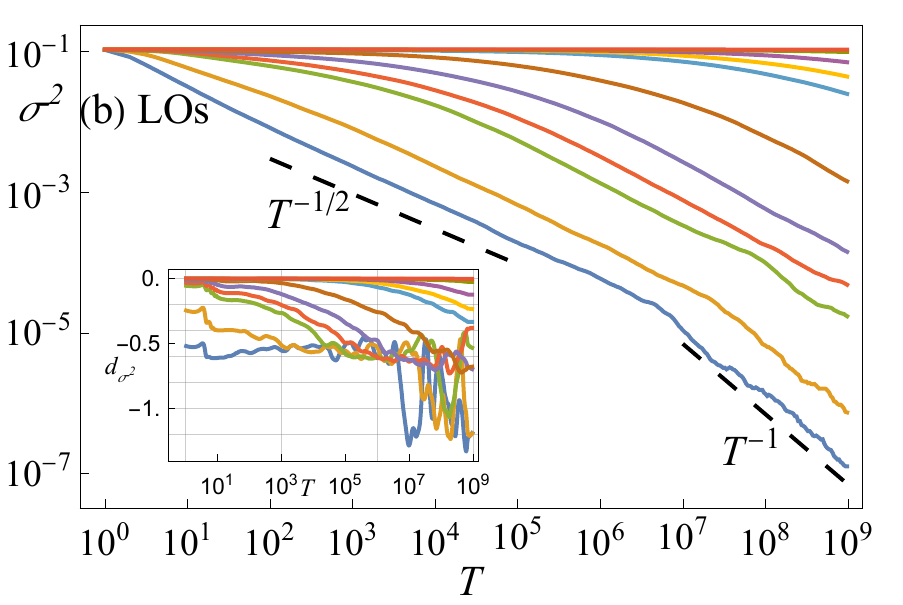}
    \caption{
    Variance $\sigma^2(T)$ of the set of time averages of EOs (a) and LOs (b) in the SRN for different $\theta$.
 Here $N=10000$ and $g=1$.
(a) Extended observables EOs with $k=\pi/2$. Three cases with $\theta=0.1,\;0.01,\;0.001$ are shown and are practically not distinguishable.
The dashed line indicates a $T^{-1}$ decay. The inset shows the local derivatives of all three curves.
(b) Local observables LOs and additional averaging over 10 trajectories (realizations).  The parameter $\theta$ takes values $\{0.5,\; 0.25,\; 0.1,\; 0.075,\; 0.05,\; 0.025,\; 0.01,\; 0.0075,\; 0.005,\; \\ 0.0025,\; 0.001\}$ from bottom blue to top red.
The two dashed lines indicate $T^{-1/2}$ and $T^{-1}$ decay respectively. 
The transition from $T^{-1/2}$ to $T^{-1}$ decay is observed for $\theta = 0.5,0.25,0.1$ (blue, orange and green curves respectively). 
The inset shows the local derivatives of all curves.}

    \label{fig4}
\end{figure}

In Fig.~\ref{fig4}(b) we show the variance $\sigma^2(T)$ for finite time averages of LOs. At times $T < T_E$ we see  approximately constant behavior $\sigma^2(T) \approx \sigma^2(0)$. For $T > T_E$ we observe the diffusive decay $\sigma(T) \sim T^{-1/2}$. For even larger averaging times $T>T_D$ we observe a transition to $\sigma^2(T) \sim T^{-1}$ due to finite size effects \cite{thudiyangal2021fragile}. Clearly the ergodization time $T_E$ grows
upon approaching the integrable limit. Together with the fast thermalization of EOs we arrive at a very good agreement of our observations with the predictions in Fig.~\ref{fig2}(b).

\begin{figure}[t]
    \centering
    \includegraphics[width=\linewidth]{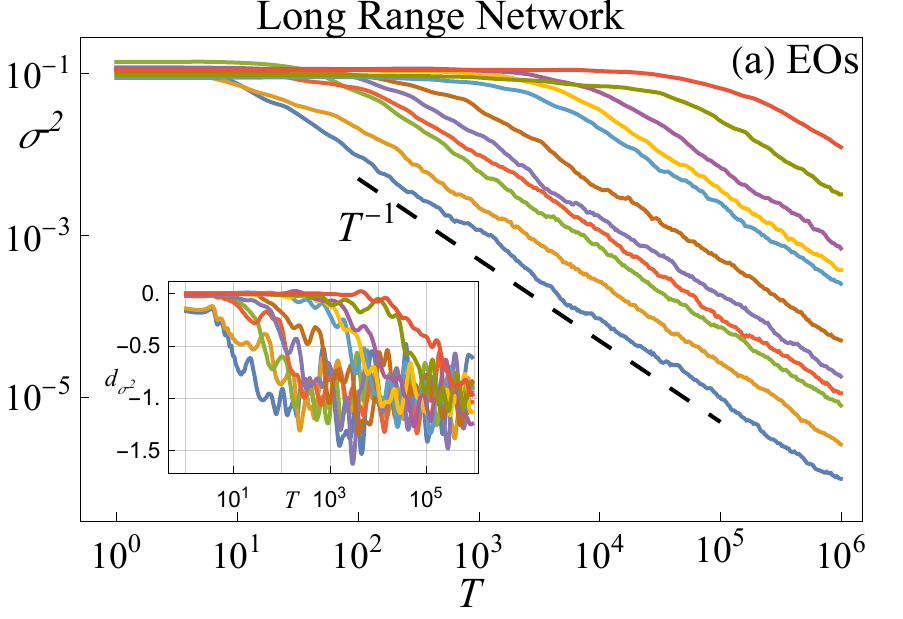}
    \includegraphics[width=\linewidth]{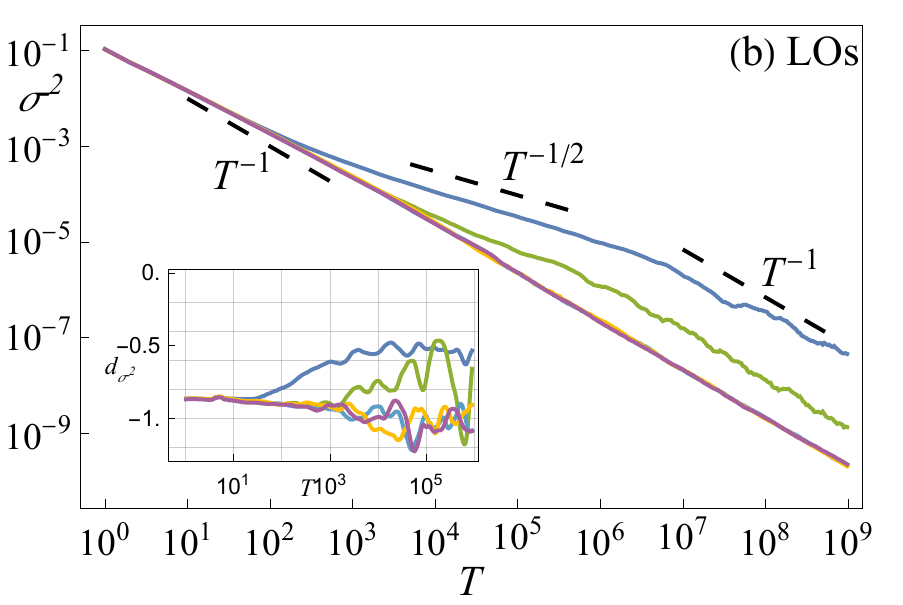}
    \caption{Variance $\sigma^2(T)$ of the set of time averages of EOs (a) and LOs (b) in the LRN for different $g$. Here $N=100000$ and $\theta=0.33\pi$.
    (a) Extended observables EOs with $k=\pi/2$. The parameter $g$ takes values $\{0.5,\; 0.25,\; 0.1,\; 0.075,\; 0.05,\; 0.025,\; 0.01,\; 0.0075,\; 0.005,\; \\ 0.0025,\; 0.001\}$ from bottom blue to top red. The dashed line indicates a $T^{-1}$ decay. The inset shows the local derivatives of all curves.
     (b) Local observables LOs and additional averaging over 10 trajectories (realizations).  Four cases with $g=0.5,\;0.1,\; 0.005\; 0.0025$.
    Three dashed lines indicate $T^{-1}$, $T^{-1/2}$ and again $T^{-1}$ decay respectively. 
    The transition from $T^{-1}$ to $T^{-1/2}$ and back to $T^{-1}$ decay is observed for $g = 0.5,\; 0.1$ (blue and green curves respectively). 
    The inset shows the local derivatives of all curves.}
    \label{fig5}
\end{figure}

\subsection{Long-range network}
In the LRN case we fix $\theta = 0.33\pi$ without loss of generality. The system size $N=10^5$. For $g = 0$ the normal mode coefficients $|c_k|$ are constant in time. The EOs are frozen, indicating another integrable limit. Upon increasing $g \neq 0$ the EOs get intercoupled with an all-to-all coupling  into a LRN. We study the statistics of local and extended observables in the LRN in close proximity to the corresponding integrable limit.

\subsubsection{Extended observables}

In Fig.~\ref{fig5}(a) we show the variance $\sigma^2(T)$ for finite time averages of EOs for various values of $g$. 
At times $T < T_E$ we see  approximately constant behavior $\sigma^2(T) \approx \sigma^2(0)$. For times $T > T_E$ we observe a
$\sigma^2(T) \sim T^{-1}$ decay. 
 Clearly the ergodization time $T_E$ grows
upon approaching the integrable limit $g \rightarrow 0$. We arrive at a very good agreement of our observations with the prediction in Fig.~\ref{fig3}(a).

\subsubsection{Local observables}

In 
Fig.~\ref{fig4}(b) we plot the variance $\sigma^2(T)$ for finite time averages of LOs for various values of $g$ approaching zero.  All curves show the same result - an immediate decay $\sigma^2 (T) \sim T^{-1}$ from the shortest averaging times on. This holds even in the integrable limit itself for $g = 0$. Similar results have been shown by \cite{vulpiani2021thermalization} -- due to the central limit theorem a random transformation of the set of action variables shows statistical properties similar to those of a mixing system when indeed it is not. Such a measurement does not imply true ergodicity, as follows from what comes next. 
We also notice a transition to $\sigma^2(T) \sim T^{-1/2}$ starting at roughly $T \approx T_E$ at which EOs start to thermalize, therefore inducing diffusion in real space. A subsequent transition to $\sigma^2(T) \sim T^{-1}$ happens for $T > T_D$ due to finite size effects.
The observed thermalization is in good agreement with the prediction in Fig~.\ref{fig3}(b).
\begin{figure}[t]
    \centering
    \includegraphics[width=\linewidth]{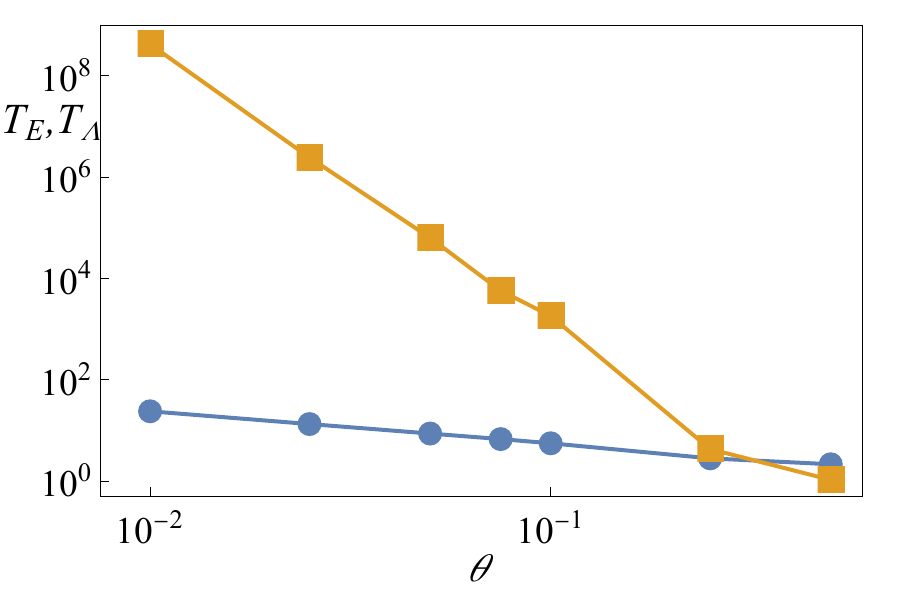}
    \caption{Short-range network case. Lyapunov time $T_\Lambda$ (blue circles) and ergodization time $T_E$ (orange rectangles) extracted from Fig.~\ref{fig4}(b) against the control parameter $\theta$. We determine $T_E$ from local derivatives as the first time when $d_{\sigma^2} = -0.25$.}
    \label{fig6}
\end{figure}

\subsection{Time scales}

\begin{figure}[t]
    \centering
    \includegraphics[width=\linewidth]{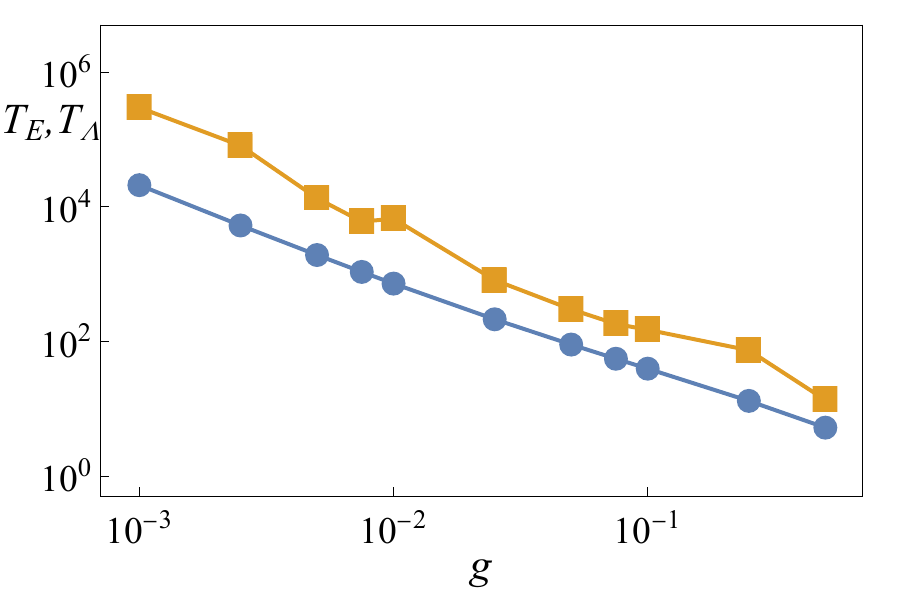}
    \caption{Long-range network case. Lyapunov time $T_\Lambda$ (blue circles) and ergodization time $T_E$ (orange rectangles) extracted from Fig.~\ref{fig5}(a) against the control parameter $g$. We obtain $T_E$ from local derivatives as the first time when $d_{\sigma^2} = -0.75$.}
    \label{fig7}
\end{figure}

In both SRN and LRN cases we compare the ergodization times $T_E$ with the characteristic Lyapunov time scales $T_\Lambda$ of chaotic dynamics. As the variance of finite time averages of actions $\sigma^2(T)$ is expected to transition from nearly constant behavior to  $\sigma^2(T)\sim T^{-1/2}$ in SRN and $\sigma^2(T)\sim T^{-1}$ in LRN we follow the local derivatives $d_{\sigma^2}$ of the variance curves $\sigma^2(T)$ , see insets of Fig.~\ref{fig4} and Fig.~\ref{fig5}. The ergodization time $T_E$ is extracted as the time when $d_{\sigma^2} = -0.25$ in SRN and $d_{\sigma^2} = -0.75$ in LRN for the first time. 
The Lyapunov times $T_\Lambda$ are computed for the same parameter values and plotted against $T_E$ in Fig.~\ref{fig6} for SRN and Fig.~\ref{fig7} for LRN. In the SRN we notice $T_E \gg T_\Lambda$ such that the ratio $T_E/T_\Lambda$ grows quickly as the system approaches integrable limit $\theta \rightarrow 0$. In the LRN case $T_\Lambda \sim T_E$ and their ratio is practically constant upon approaching the integrable limit $g \rightarrow 0$.

\subsection{Lyapunov Spectra and Tangent Vectors} 

\begin{figure}[t!]
    \centering
    \includegraphics[width=\linewidth]{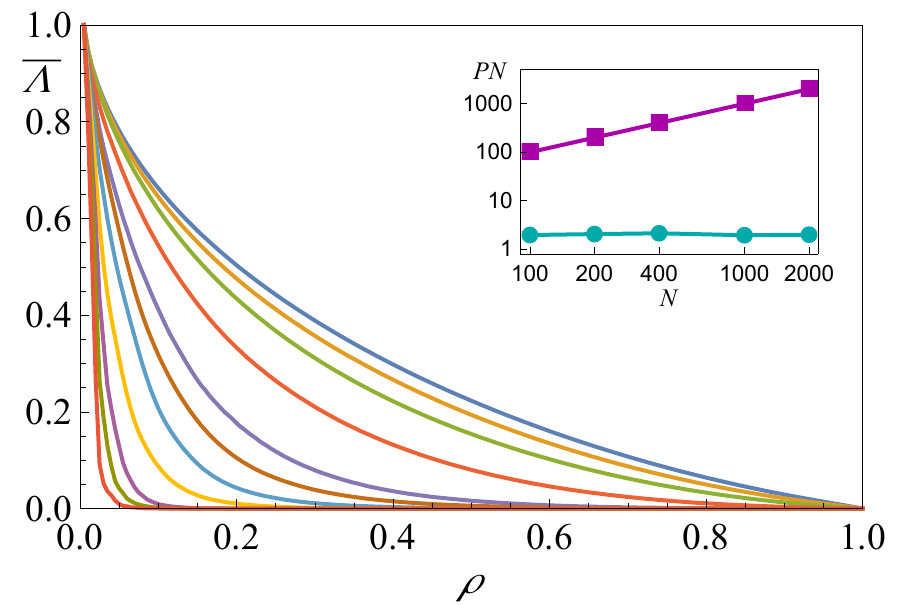}
    \caption{The rescaled Lyapunov spectra $\bar\Lambda (\rho)$ in SRN regime. Parameters are $N = 200$, $g=1$. Different curves correspond to
$\theta = \{0.1, 0.075, 0.05, 0.025, 0.01, 0.0075, 0.005, 0.0025, 0.001,\\ 0.00075, 0.0005\}$ from top to bottom. Data are used from Ref.~\cite{malishava2022lyapunov}.
Inset: $PN$ of tangent vector versus $N$ for $\theta=0.001$. 
Cyan circles: LO representation. Magenta squares: EO representation.}
    \label{fig8}
\end{figure}

\begin{figure}[t!]
    \centering
    \includegraphics[width=\linewidth]{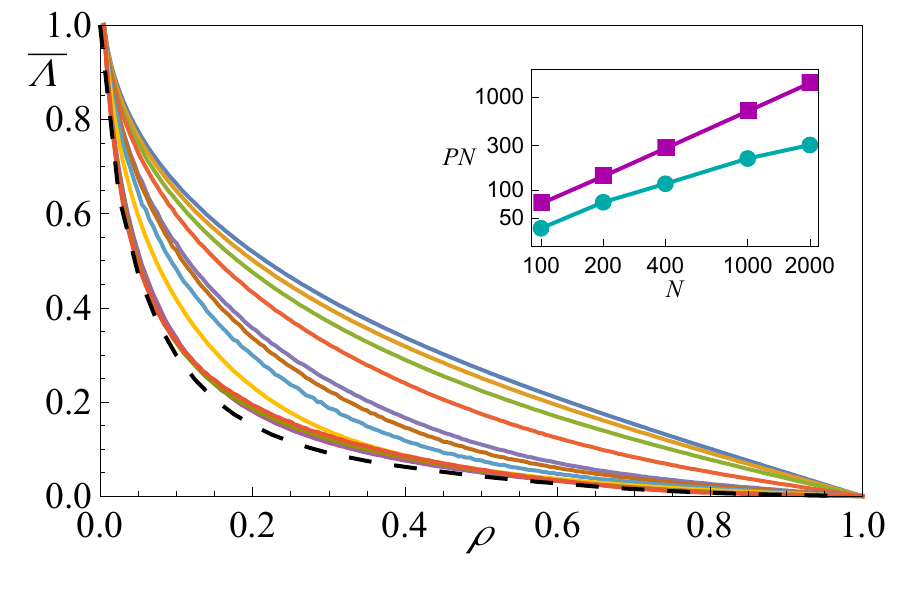}
    \caption{The rescaled Lyapunov spectra $\bar\Lambda (\rho)$ in LRN regime. Parameters are $N = 200$, $\theta=0.33$. Different curves correspond to
$g= \{0.1, 0.075, 0.05, 0.025, 0.01, 0.0075, 0.005, 0.0025, 0.001, \\0.00075, 0.0005\}$ from top to bottom. Data are used from Ref.~\cite{malishava2022lyapunov}.
Inset: $PN$ of tangent vector versus $N$ for $g=0.001$. 
Cyan circles: LO representation. Magenta squares: EO representation.}
    \label{fig9}
\end{figure}

In Fig.~\ref{fig8} and Fig.~\ref{fig9} we show the rescaled Lyapunov spectra $\bar\Lambda(\rho) = \Lambda(\rho)/\Lambda_{max}$ corresponding to the SRN and the LRN respectively. 

The rescaled spectrum tends to a non-analytic function in the SRN upon approaching the integrable limit in Fig.~\ref{fig8}.
Close to the limit the spectrum shows exponential decay with a (length scale) exponent which diverges at the very integrable limit \cite{malishava2022lyapunov}. The slowing down of thermalization is characterized by two diverging scales - a time scale and a length scale.

The rescaled spectrum tends to an analytic function in the LRN upon approaching the integrable limit in Fig.~\ref{fig9}.
The slowing down of thermalization is characterized by one diverging time scale only.

In the insets of Fig.~\ref{fig8} and Fig.~\ref{fig9} we plot the participation $PN$ number of the tangent vector corresponding to the largest Lyapunov exponent as a function of system size for. For the SRN case the parameters are $g=1,\; \theta=0.001$. For the LRN case the parameters are 
$\theta=0.33\pi,\;g=0.001$. The participation numbers depend on the basis choice of the tangent vector. Thus we compute the $PN$ in both LO and EO representations.

In the SRN case (inset in Fig.~\ref{fig8}) the tangent vectors in LO representation are localized due to rare chaotic resonances in real space. Therefore the
$PN$ number in LO representation is predicted to be system size independent.
Since a localized distribution in real space turns delocalized in reciprocal space, the EO representation results in a $PN$ which grows and scales with the system size.

In the LRN case (inset in Fig.~\ref{fig9}) resonances are expected to appear almost everywhere in the EO representation.Thus we expect the $PN$ number
in EO representation to grow and scale with the system size. This is in accord with our numerical observation.  
Interestingly the delocalized nature of the tangent vector in the EO representation does not imply that the LO representation will result in localized structures.
Indeed the numerical computation show that the $PN$ number in LO representation also grows and scales with the system size.

\section{Discussion and Conclusion\label{Sec6}} 

One of our main findings is that the choice of observables for a thermalization study can lead to ambiguous conclusions. In particular, we consider macroscopic systems which are tuned close to an integrable limit. Their slow adiabatic invariants are given by the conserved actions at the very integrable limit. Making these actions the observables of choice will obviously result in the correct thermalization analysis. We study two different classes of
weakly nonintegrable lattice systems with discrete translational invariance, in which the actions are interacting through either a long range network, or a short range network. For LRNs the actions are extended in real space, so the relevant observables are extended. For SRNs the actions are local in real space, thus the relevant observables are local. As we show in particular, the choice of the 'wrong' observables - LOs for LRNs or EOs for SRNs - will result in a seemingly quick thermalization without any slowing down upon approaching the integrable limit, including the limit itself. At the same time the correctly chosen observables - EOs for LRNs and LOs for SRNs - will show a dramatic slowing down of thermalization. We also show that a simultaneous study of both LO and EO thermalization allows to unambiguously identify the slowing down, and even conclude which class - LRN or SRN - is under study.


As a result of our study, we extract the ergodization time scale $T_E$ as a function of the control parameter which tunes the distance from the integrable limit. 
We also compute the largest Lyapunov exponent and its inverse - the Lyapunov time $T_\Lambda$. It follows that the LRN class is characterized by only one
time scale as
$T_E\sim T_\Lambda$. 
At variance to the above, the SRN class must be characterized by other diverging scales, as we observe that 
$T_E\gg T_\Lambda$ in accord with previous observations for Josephson junction arrays \cite{mithun2019dynamical} and Klein-Gordon chains \cite{danieli2019dynamical}. This second time or length scale was predicted to arise from low densities of rare resonances in real space, and the need for
these resonances to migrate over the increasing and diverging average distance between them \cite{mithun2019dynamical}.

For times larger than the ergodization time scale $T_E$ the finite time average distributions of LOs will show a diffusive convergence of their variance 
$\sigma^2 \sim T^{-1/2}$. The impact of a finite size $N$ of the system results in a diffusion time scale $T_D\sim N^2$ when the variance crosses over from to $\sigma^2 \sim T^{-1}$. These findings are in line with studies on Hamiltonian dynamics such as Josephson junctions \cite{thudiyangal2021fragile}.
The above scheme of identifying the correct network class relies on measuring both LOs and EOs and on varying the control parameter of the distance to the integrable limit. Interestingly we can tell the right network class also if we do not vary that control parameter, but instead vary the system size $N$.
For that we note that the computation of the largest Lyapunov exponent comes with its corresponding tangent vector information. We use this information to 
compute its average participation number $PN$ in both the direct local space, and in reciprocal space. For the SRN we already expect the tangent vector to be highly local in real space, thus delocalized in reciprocal space. Indeed, $PN$ is essentially independent of $N$ in direct space, but scales $PN \sim N$ in reciprocal space. At variance to that, the LRN results in a $PN \sim N$ scaling for both spaces. Therefore we can tell the network class from a finite size analysis of the participation number of the tangent vector, without varying the distance to the integrable limit.


In a recent study we computed the entire Lyapunov spectrum and analyzed its scaling properties upon varying the control parameter of the distance to the integrable limit \cite{malishava2022lyapunov}. For LRNs the rescaled Lyapunov spectra converge to an analytic function, leaving us with only one characteristic time scale $T_\Lambda$ which is close to $T_E$. At variance, for SRN the rescaled Lyapunov spectrum converges to a non-analytic function, and on that route one can extract a second divering length scale \cite{malishava2022lyapunov}, in agreement with this work and previous studies.
An advantage of Lyapunov spectra computation is that they are independent of any coordinate basis choice.  Nonzero Lyapunov spectra necessarily imply chaotic \textit{mixing} dynamics. On the downside we note that while the largest LCE can be experimentally assessed, the entire Lyapunov spectrum appears to be not experimentally measurable.

One of the most interesting open questions is whether there are other network classes which have experimental relevance. This can happen for lattice systems in the limit of weak coupling with algebraic coupling decay along the lattice. Another interesting case concerns weak two-body interactions and thus weak nonlinearities in the presence of disorder, such that the actions at the integrable limit are either Anderson localized or extended but fractal or multi-fractal. From a more mathematical perspective it would be interesting to study weak perturbations of such known integrable systems as the Toda chain \cite{toda1967vibration} or the Ablowitz-Ladik system \cite{ablowitz1976nonlinear}. Last but not least all studies need to done also for higher lattice dimensionality.

\textit{Acknowledgments:}
The authors thank Carlo Danieli, Ihor Vakulchyk for valuable discussions and Boris Fine for pointing our attention to the tangent vector information.
This work was supported by the Institute for Basic Science (Project number: IBS-R024-D1).

\appendix
\section{Computing Lyapunov times\label{appendix:Lyapunov times}}

To compute Lyapunov exponents we follow the prescription given in \cite{benettin1980lyapunov}. We introduce an orthogonal set of vectors $\lbrace \vec{w}_i \rbrace$ as a deviation from some unperturbed trajectory $\vec x$.
\begin{equation}
\label{ape_eq_1}
    \vec\Psi_i(t) = \vec x(t) + \vec w_i(t)
\end{equation}

\begin{figure}
\centering
\begin{minipage}[b]{.45\textwidth}
\includegraphics[width=\textwidth]{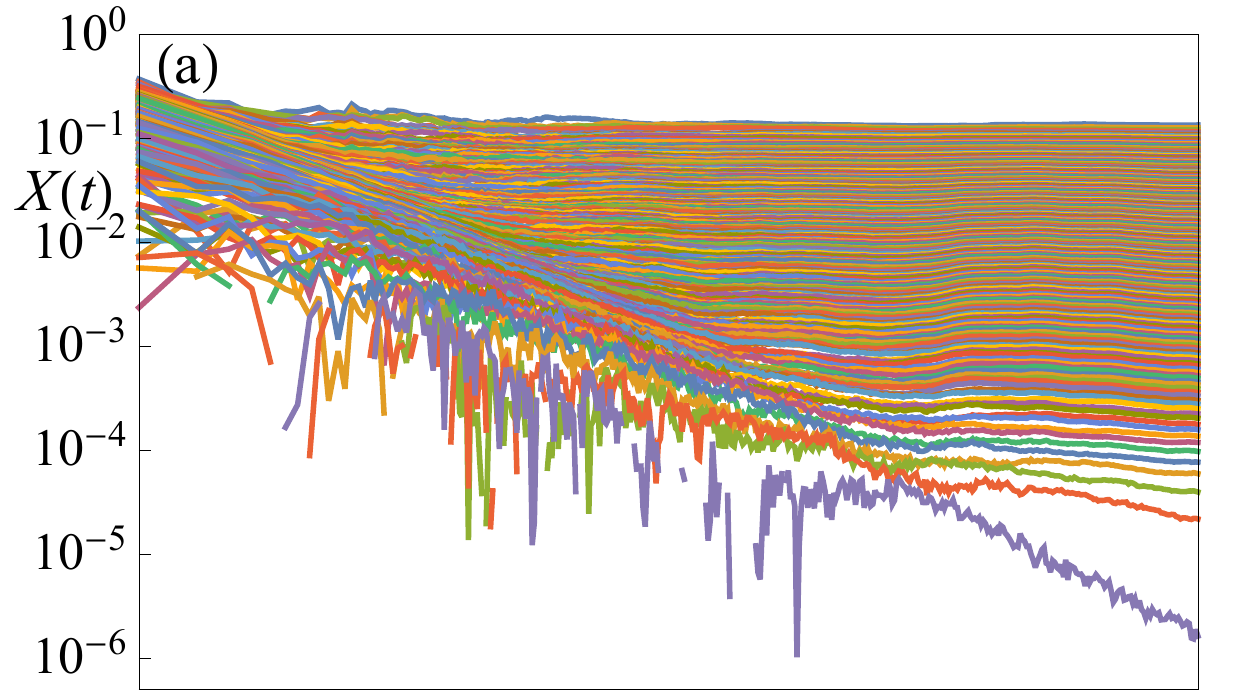}
\end{minipage}\hfill
\begin{minipage}[b]{.45\textwidth}
\includegraphics[width=\textwidth]{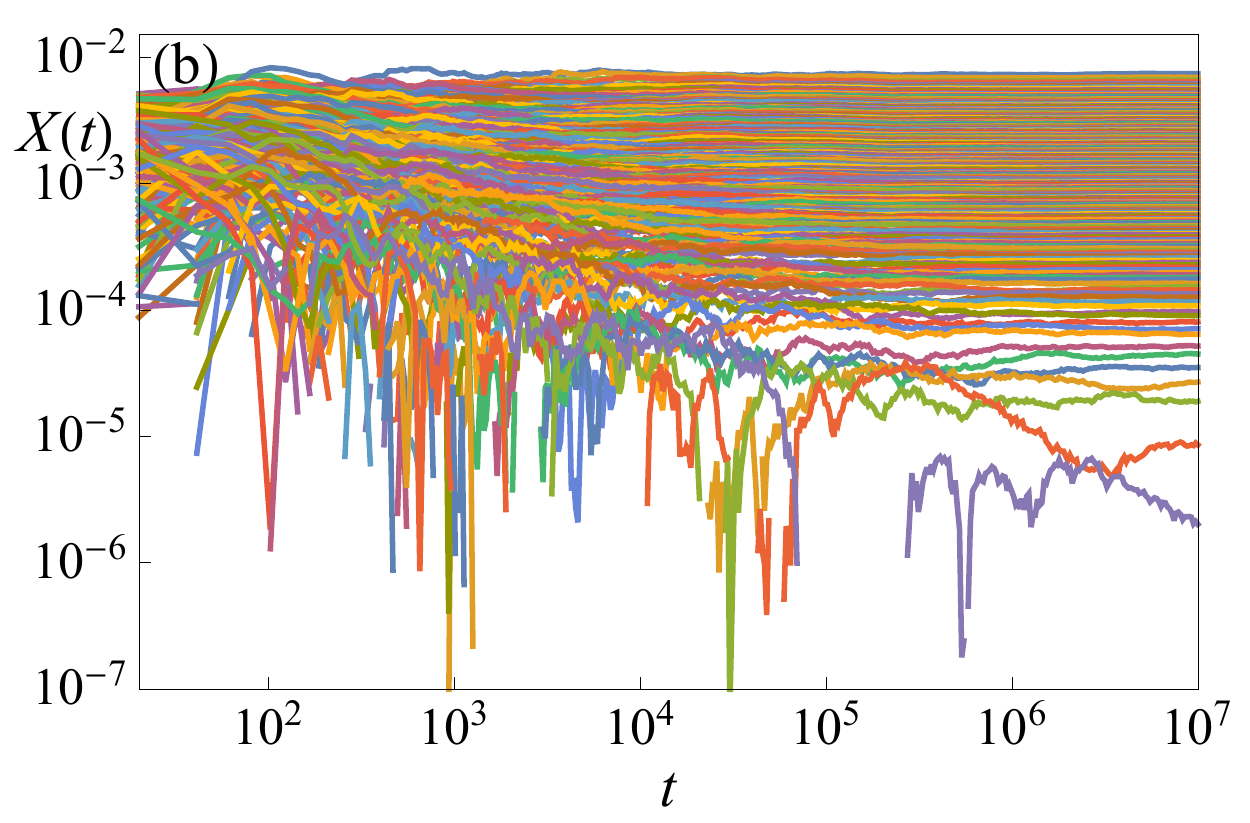}
\end{minipage}
\caption{The evolution of positive transient Lyapunov exponents. a) SRN case with angle $\theta = 0.1$ and nonlinearity $g = 1.0$, b) LRN case with angle $\theta = 0.33\pi$ and $g = 0.1$. For both cases system size $N = 200$.\label{ape:fig1}}
\end{figure}

The direction of each vector $\{ w_i \}$ corresponds to a direction of an exponential growth or contraction of the distance between unperturbed trajectory $\vec x$ and the perturbed $\vec \Psi$. The evolution of tangent vectors is performed using the corresponding equations of motion derived below. We measure the magnitude of growth  $\gamma(t) = |\vec w(t)|$ of each tangent vector and compute transient Lyapunov exponents $X_i(t) = 1/t\sum_\tau^t \log \gamma(\tau)$ after which the tangent vectors are orthonormalized using a Gram-Schmidt procedure. The evolution of positive transient Lyapunov exponents $X(t)$ is shown in Fig.~\ref{ape:fig1}. After an initial decay the transient Lyapunov exponents saturate. The saturated values are taken as final values for Lyapunov exponents $\Lambda$. Due to the conservation of the norm two exponents are expected to attain zero value. In the figure we see one of them (bottom most purple line) tending to zero with increasing time and no saturation. 

Before deriving equations for deviation vectors $\vec w_i$ we first define the linear part of evolution : 
\begin{eqnarray}
\label{ape_eq_2}
&& \alpha_n^{A}[\vec\Psi(t)] \equiv \cos^2\theta\psi_n^A(t)- \cos\theta\sin\theta\psi_{n-1}^B(t)\nonumber  \\
&&+ \sin^2\theta\psi^A_{n +1}(t) + \cos\theta\sin\theta\psi_n^B(t) \nonumber \\
\nonumber\\
&& \alpha_n^{B}[\vec\Psi(t)] \equiv \sin^2\theta\psi_{n - 1}^B(t)- \cos\theta\sin\theta\psi_{n}^A(t) \nonumber  \\
&&+ \cos^2\theta\psi^B_{n}(t) + \cos\theta\sin\theta\psi_{n + 1}^A(t) \;.
\end{eqnarray}
We start from the nonlinear EoM Eq.~\eqref{eomsNonLinear} and substitute in Eq.~\eqref{ape_eq_2}:
\begin{eqnarray}
&& \psi_n^{A}(t + 1) = e^{ i g |\alpha^A_n[\vec{x}(t)+\vec{w}(t)]|^2} \alpha^A_n\left[(\vec{x}(t)+\vec{w}(t))\right] \nonumber \\
&& \psi_n^{B}(t + 1) = e^{ i g |\alpha^B_n[\vec{x}(t)+\vec{w}(t)]|^2} \alpha^B_n\left[(\vec{x}(t)+\vec{w}(t))\right]. \nonumber \\
\label{ape_eq_3}
\end{eqnarray}
Expanding the nonlinear term and keeping terms only in the $1$st order of $\vec{w}$ results in
\begin{eqnarray}
\label{ape_eq_4}
   && |\alpha^p_n[\vec{x}(t)+\vec{w}(t)]|^2 = |\alpha^p_n[\vec{x}(t)]+\alpha^p_n[\vec{w}(t)]|^2 = \nonumber \\ &&\alpha^p_n[\vec{x}(t)]\alpha^p_n[\vec{x}(t)]^* + \alpha^p_n[\vec{w}(t)][\alpha^p_n[\vec{w}(t)]^* + \nonumber \\ &&\alpha^p_n[\vec{w}(t)]\alpha^p_n[\vec{x}(t)]^* + \alpha^p_n[\vec{x}(t)]\alpha^p_n[\vec{w}(t)]^* \approx  \nonumber \\
   && |\alpha^p_n[\vec{x}(t)]|^2 + \Delta^p_n(t),
\end{eqnarray}
where
\begin{eqnarray}
   &&\Delta^p_n(t) = \alpha_n^p[\vec{x}(t)]\alpha_n^p[\vec{w}(t)]^* + c.c. \nonumber \\
\end{eqnarray}
Thus we can rewrite the exponential term in Eq.~\eqref{ape_eq_3}:
\begin{eqnarray}
   e^{ i g |\alpha_n^p[\vec{x}(t)+\vec{\varepsilon}(t)]|^2} = e^{ i g |\alpha_n^p[\vec{x}(t)]|^2}\left[1 + i g \Delta_n^p(t)\right], 
\end{eqnarray}
and using the linearity of $\alpha_n^p[\vec{\Psi(t)}]$ we finally arrive at the
following linear equations:
\begin{eqnarray}
   &&w^p(t+1) = e^{ i g |\alpha_n^p[\vec{x}(t)]|^2}\Big\lbrace \alpha_n^p[\vec{w}(t)] + i g \Delta_n^p(t)\alpha_n^p[\vec{x}(t)]\Big\rbrace. \nonumber \\
\end{eqnarray}

\bibliography{bibliography}

\end{document}